\documentclass[conference]{IEEEtran}
\usepackage{cite}
\usepackage{amsmath,amssymb,amsfonts}
\usepackage{algorithmic}
\usepackage{graphicx}
\usepackage{textcomp}
\usepackage{booktabs} 
\usepackage{multirow} 
\usepackage{xcolor}
\usepackage{balance}
\def\BibTeX{{\rm B\kern-.05em{\sc i\kern-.025em b}\kern-.08em
    T\kern-.1667em\lower.7ex\hbox{E}\kern-.125emX}}

\begin{document}

\title{An Analysis of Untrained Deep Reservoir Networks for Audio Surveillance\\

}
\ifx\anonymize\undefined
\author{\IEEEauthorblockN{Corrado Baccheschi}
\IEEEauthorblockA{\textit{Department of Computer Science} \\
\textit{University of Pisa}\\
Pisa, Italy \\
(ORCID: 0009-0002-6500-9578) }

\and
\IEEEauthorblockN{Patrizio Dazzi}
\IEEEauthorblockA{\textit{Department of Computer Science} \\
\textit{University of Pisa}\\
Pisa, Italy \\
(ORCID: 0000-0001-8504-1503)}

}

\else
\author{Anonymous submission}
\fi

\maketitle

\begin{abstract}
In this paper, we investigate untrained recurrent models from the Reservoir Computing (RC) paradigm for audio surveillance, focusing on bidirectional Echo State Networks with different depths, from shallow to deep configurations, for emergency sound event detection. We evaluate these models on the MIVIA Audio Events dataset in a multiclass setting across different Signal-to-Noise Ratio (SNR) levels, with the goal of assessing the trade-off between depth, recognition performance, and computational efficiency. We compare the proposed architectures against fully trained recurrent and convolutional-recurrent baselines, namely Bidirectional Long Short-Term Memory networks (BiLSTMs) and Convolutional Recurrent Neural Networks (CRNNs). Results show that deep and shallow reservoir-based models achieve competitive recognition rates, with deeper variants being more robust in highly noisy conditions and shallower ones offering the most favorable efficiency profile, particularly on edge devices such as the NVIDIA Orin. In addition, the proposed approach remains robust across different input representations, including log-Mel spectrograms and MFCCs with varying resolutions. These findings highlight untrained reservoir architectures as a promising solution for resource-constrained audio surveillance scenarios.
\end{abstract}


\section{Introduction}
In the realm of audio classification, audio surveillance aims to identify critical emergency events within an audio stream, ranging from environmental sounds (e.g., glass breaking, explosions) to human-related events (e.g., screams, falls). To design such systems, a common approach is to transform raw audio signals into feature representations that can be effectively processed by machine learning (ML) models. In this regard, hand-crafted features such as Mel-Frequency Cepstral Coefficients (MFCCs) and Mel spectrograms are widely adopted as inputs for traditional ML methods (e.g., SVM, KNN) and have demonstrated strong performance across various applications~\cite{systematicrev}. However, audio surveillance systems pose significant challenges: they must achieve high recognition accuracy,  and, most importantly, they have to operate with low latency, even in resource-constrained devices~\cite{colangelo2017}\cite{greco2021}\cite{dazzi2025internet}. 
Deep learning models, including recurrent and convolutional architectures, have further improved performance by effectively capturing temporal and local patterns through hierarchical representations~\cite{systematicrev}\cite{dlreview}.
Pretrained audio-tagging CNNs, such as PANNs~\cite{PANNS}, as well as self-supervised models based on Transformer architectures, such as Audio Spectrogram Transformer (AST)~\cite{gong21b_interspeech}, have also been employed for audio tasks, often further refined through large-scale pretraining~\cite{gong_psla}. However, despite their strong performance, these models typically come with high computational costs, limiting their applicability in resource-constrained environments and raising concerns about their efficiency~\cite{dlreview}. In addition, recurrent architectures are notoriously difficult to train, as they suffer from vanishing gradients~\cite{gradientissues}, which hinder the modeling of long-term dependencies and reduce their suitability for real-world deployment scenarios. Motivated by these challenges, we explore the trade-off between performance and efficiency offered by the Reservoir Computing (RC) paradigm~\cite{LUKOSEVICIUS2009127} \cite{verstraeten2007experimental}, which provides a framework for designing RNNs where the recurrent dynamics are not trained; while only a final readout layer is learned. In fact, by keeping the recurrent component fixed after random initialization, RC is able to mitigate gradient-related optimization issues while retaining the ability to model temporal dependencies. Thus, in this work, we investigate whether deep recurrent models from the RC paradigm can provide an effective and robust alternative for audio surveillance in noisy and resource-constrained environments. To this end, we adopt Deep Echo State Networks (DeepESNs)~\cite{GALLICCHIO201787}, a layered extension of the Echo State Network (ESN)~\cite{doi:10.1126/science.1091277}\cite{Herbert}, the most widely used architecture in RC. We focus on a bidirectional variant of DeepESN, as bidirectionality has become a standard design choice in audio classification due to its ability to capture temporal dependencies more effectively~\cite{colangelo2017} \cite{systematicrev}.
We then pose the following research questions:
\begin{description}
    \item[RQ1] How does DeepBiESN perform on the MIVIA audio surveillance dataset compared to fully-trained approaches?
    \item[RQ2] How does DeepBiESN compare in terms of efficiency on CPU and GPU against fully trained models? Does a trade-off exist between the efficiency of the model and accuracy?
     \item[RQ3] How robust is DeepBiESN to the changes in feature representations with respect to fully trained architectures?
\end{description}

In summary, our objective is not only to assess whether reservoir-based bidirectional recurrent models can be effectively applied to audio surveillance, but also to understand how architectural depth affects robustness, efficiency, and overall deployment suitability. In this perspective, comparing shallow and deep reservoir configurations is essential to identify the most appropriate operating regime for resource-constrained and low-latency scenarios. To this end, we first evaluate different reservoir configurations in a server-based setting, and subsequently extend the analysis to edge environments by testing the proposed approach on an NVIDIA Orin platform. 
This work is organized as follows: Section~\ref{rel} exposes similar works in the field; Section~\ref{meth} explains our methodology and the chosen approaches; Section~\ref{experiments} reports our experiments on the MIVIA audio events dataset comparing with the previous results in the literature; Section~\ref{robustness} analyses the robustness of DeepBiESN in the changes of feature representation; finally, Section~\ref{conclusions} summarizes the findings and suggests potential future directions.

\section{Related work}
\label{rel}
Early work in audio analysis relied on traditional classifiers, including Support Vector Machines (SVM)~\cite{svm} and k-Nearest Neighbors (KNN), often achieving solid performance when paired with carefully engineered features~\cite{systematicrev}. Nowadays, with the rise of deep learning, more complex architectures have been introduced to automatically learn hierarchical representations from audio data. In this context, convolutional (CNNs) and recurrent neural networks (RNNs) have certainly shown strong performance by effectively capturing local and temporal patterns in audio data~\cite{dlreview}. Audio surveillance can be seen as a specific subtask of audio classification, focusing on the detection of emergency events, whether environmental or human-related. This problem has been studied across a variety of scenarios. For example, Foggia et al.~\cite{Foggia2015} introduced the MIVIA Audio Events dataset and proposed a Bag-of-Words approach with SVM for detecting glass breaking, gunshots, and screams. Subsequent works have applied RNNs~\cite{colangelo2017} and convolutional-recurrent neural networks (CRNN)~\cite{greco2021} to the same benchmark, progressively improving recognition performance at the cost of increased model complexity and training requirements. Beyond this dataset, audio detection has also been investigated in fire and burning scenarios for scream identification~\cite{burningscream}, in construction site environments for detecting hazards and worker-related sounds~\cite{constructionsites}, and in safety systems targeting vulnerable populations such as women and children~\cite{womenchild}. As an alternative to fully trained architectures, Reservoir Computing (RC) models have been applied to a wide range of audio-related tasks. For instance, Scarpiniti et al.~\cite{Scarpiniti2023} employed a Leaky Echo State Network (LeakyESN) for classifying vehicle and tool sounds in construction site scenarios, demonstrating competitive performance with significantly reduced training complexity compared to CNN. Echo State Networks have also been successfully applied to real-time audio processing~\cite{esnaudio}, music-related tasks~\cite{esnmusic}\cite{onsetoff}, and speech processing~\cite{esnspeech}, showing their suitability for modeling audio tasks. However, despite these advances, the use of \emph{deep} reservoir architectures in audio surveillance remains largely unexplored. To the best of our knowledge, this is the first systematic analysis of deep untrained reservoir models in noisy audio surveillance.
\section{Proposed approach}
\label{meth}
In this Section, we explain the methodology and the approaches adopted for the work presented.

\subsection{Reservoir computing}
Reservoir computing (RC)~\cite{LUKOSEVICIUS2009127,verstraeten2007experimental} avoids training the recurrent dynamics and instead keeps the input and recurrent weights fixed after random initialization, while learning only a final linear layer called $readout$, which is typically implemented as a ridge regression in closed-form. Among the various models within the RC paradigm, the Echo State Network (ESN)~\cite{doi:10.1126/science.1091277,Herbert} is the most widely adopted architecture.

In a deep learning context, the Deep Echo State Network (DeepESN)~\cite{GALLICCHIO201787}\cite{Gallicchio2018} is a specific variant of an ESN model that leverages a layered architecture. DeepESN encodes the input sequence $$\mathbf{x}(t) \in \mathbb{R}^{N_X}$$ as a hierarchy of representations $$\mathbf{h}^{(l)}(t) \in \mathbb{R}^{N_R}~\text{with}~l = 1, \dots, L$$ Let $\mathbf{h}^{(l)}(t)$ the computed state at the time $t$ for a layer $l$, the state transition function of a DeepESN layer is
\begin{multline}
\label{1}
    \mathbf{h}^{(l)}(t) = (1-a)\, \mathbf{h}^{(l)}(t-1)  \\
    + a  \tanh \left(\mathbf{W_h}^{(l)}  \mathbf{h}^{(l)}(t-1) + \mathbf{W_x}^{(l)} \mathbf{x}^{(l)}(t) + \mathbf{b}^{(l)} \right)
\end{multline}
where the input weights $$\mathbf{W_x} \in \mathbb{R}^{N_R \times N_X}$$ and bias $$\mathbf{b} \in \mathbb{R}^{N_R}$$ are randomly initialized, recurrent weights $$\mathbf{W_h} \in \mathbb{R}^{N_R \times N_R}$$ are randomly initialized and rescaled to satisfy the echo state property \cite{Herbert}, and  $0 < a \leq 1$ is the leakage constant~\cite{Jaeger2007} which handles the influence of the past inputs. To introduce bidirectionality in a DeepESN, we encode the input sequence in both forward and backward temporal directions as in ~\cite{bianchi2018bidirectionaldeepreadoutechostate}, and then concatenate them as $$\bar{\mathbf{h}}^{(l)}(t) = \left[\begin{smallmatrix}\overrightarrow{\mathbf{h}}^{(l)}(t) & \overleftarrow{\mathbf{h}}^{(l)}(t)\end{smallmatrix}\right]$$ We refer to this particular architecture as \textit{DeepBiESN} where the input to layer $l > 1$ is  $$\mathbf{x}^{(l)}(t) = \bar{\mathbf{h}}^{(l-1)}(t)$$ All the code has been released for research and validation purposes\footnote{GitHub: https://github.com/Bakko000/TorchDeepESN/}.

\subsection{Audio feature representations}
\label{inputrepr}
Audio signals are represented as log-scaled Mel spectrograms, a widely adopted input representation for audio classification tasks~\cite{systematicrev}\cite{logmelspectrogram}\cite{logmelspectrogram2}. Each waveform is then converted to a Mel spectrogram using a Short-Time Fourier Transform (STFT) with window size $N_{\text{FFT}} = 2048$, hop length $H = 2560$, and $M = 128$ Mel filterbanks, yielding a time resolution of approximately $80$ ms per frame. The resulting spectrogram is finally converted to decibel scale and normalized per-clip to zero mean and unit variance. 

\section{Experiments}
\label{experiments}

In this Section, we detail our experimental setup to answer RQ1 and RQ2, presenting the results comparing DeepBiESN with fully-trained models.

\subsection{Experimental setup}
All experiments were conducted on a server equipped with dual CPUs (64 cores each, 128 cores total), 1.5 TB of RAM, and an NVIDIA L40S GPU with 45 GB of VRAM. Additionally, we evaluated the efficiency of our approach on an edge platform, namely the NVIDIA Jetson AGX Orin running JetPack 6, featuring a 12-core ARMv8 CPU, 64 GB of RAM, and an integrated NVIDIA Ampere GPU. As inputs to the Deep-BiESN, we feed each data point as a sequence of standardized log-mel spectrogram representations, as already presented in Section~\ref{inputrepr}.
The output layer ($readout$) is trained using an efficient closed-form ridge regression solution~\cite{Zhang2017}, and the recurrent weight matrix $\mathbf{W_h}$  for the DeepBiESNs is initialized using the fast spectral scaling strategy proposed in~\cite{Gallicchio2020fast}. For model selection, we adopt a hold-out validation strategy, allocating 15\% of the available data to a dedicated validation split. We tune DeepBiESN through an exhaustive grid search, optimizing $F1$ macro as a metric of evaluation. 
The best performing DeepBiESN model after the selection includes $L=5$, $\rho=0.5$, $a=0.5$, and $6.105\times 10^{-6}$ as the regularization coefficient, scoring $F1= 98\%$  on the validation set. We then evaluate our models on the test set following the event-based evaluation protocol proposed in~\cite{Foggia2015}
where the evaluation metric is the Recognition Rate (RR) computed as $\text{True Positives}(TP) / N$ where $N$ denotes the total number of annotated events.
To ensure a fair comparison, all models were evaluated on the same train, validation, and test partitions, using the same input representations for each experimental setting. In addition, we explored configurations of DeepBiESNs with varying network depth, retaining the best-performing models with $L=3$ and $L=1$, in order to isolate the contribution of depth to recognition performance and efficiency. This design allows us to interpret the shallow variant ($L=1$) not only as a lightweight baseline, but also as an ablation on the hierarchical structure of the proposed architecture.


\subsection{Dataset}
As a source of data, we use the MIVIA Audio Events dataset, a widely adopted benchmark for audio surveillance~\cite{Foggia2015}. It contains clips of approximately $180$ seconds each, recorded at different SNR levels, enabling the evaluation of model robustness under varying noise conditions and simulating real-world microphone distances. The dataset includes three types of events—glass breaking (GB), screams (S), and gunshots (GS)—all related to emergency scenarios. These events are superimposed on background noise (BG), increasing the difficulty of the classification task. Table~\ref{tab:mivia_dataset} reports the detailed dataset used for the experiments.

\begin{table}[htbp]
\centering
\small
\caption{Composition of the MIVIA Audio Events dataset. Number of events and total duration (in seconds) for Background  (BG), Glass Breaking (GB), Gunshots (GS), and Screams (S).}
\label{tab:mivia_dataset}
\begin{tabular}{lcc|cc}
\toprule
 & \multicolumn{2}{c}{\textbf{Training set}} & \multicolumn{2}{c}{\textbf{Test set}} \\
\cmidrule(lr){2-3} \cmidrule(lr){4-5}
\textbf{Type} & \textbf{Events} & \textbf{Duration (s)} & \textbf{Events} & \textbf{Duration (s)} \\
\midrule
BG & --   & 58{,}372 & --   & 25{,}037 \\
GB & 4{,}200 & 6{,}025 & 1{,}800 & 2{,}562 \\
GS  & 4{,}200 & 1{,}884 & 1{,}800 & 744 \\
S  & 4{,}200 & 5{,}489 & 1{,}800 & 2{,}445 \\
\bottomrule
\end{tabular}
\end{table}

\subsection{Results and comparison with the literature}
We report the test scores for the best configuration of DeepBiESN in Table~\ref{tab:comparison} for a contextual comparison with previously reported results on the MIVIA Audio Events dataset. Since preprocessing pipelines, and evaluation protocols may differ across studies, the reported recognition rates should be interpreted as a literature reference rather than as a strict ranking.  

\begin{table}[t]
\centering
\footnotesize
\setlength{\tabcolsep}{3.2pt}
\caption{Recognition rates (RR) event-based, from different studies and protocols in literature, for the MIVIA Audio Events dataset. When available, RR at different SNR levels is also included. Best in bold, second best underlined.}
\label{tab:comparison}
\begin{tabular}{lccccccc}
\toprule
\textbf{Method} & \textbf{RR} & \textbf{5dB} & \textbf{10dB} & \textbf{15dB} & \textbf{20dB} & \textbf{25dB} & \textbf{30dB} \\
\midrule
HF+BoW+SVM~\cite{Foggia2015} & 86.7 & 81.1 & 85.0 & 87.0 & 88.4 & 88.7 & 90.0 \\
HRNN~\cite{colangelo2017} & 96.5 & 90.7 & 92.4 & \underline{98.5} & 98.7 & 99.1 & \textbf{99.9} \\
DeNet~\cite{greco2021} & 97.5 & \underline{92.1} & 96.9 & 98.4 & 98.9 & 99.2 & 99.2 \\
COPE~\cite{COPE} & 96.0 & -- & -- & -- & -- & -- & -- \\
AreN~\cite{AreN} & \textbf{99.4} & -- & -- & -- & -- & -- & -- \\
SincNet-based~\cite{sincnetbengio,sincnet} & 97.1 & 87.3 & \underline{97.6} & \textbf{99.2} & \underline{99.4} & \underline{99.4} & \underline{99.4} \\
Haar~\cite{haar} & 88.6 & -- & -- & -- & -- & -- & -- \\
DeepBiESN (ours) & \underline{98.3} & \textbf{92.9} & \textbf{97.8} & \textbf{99.2} & \textbf{99.9} & \textbf{99.9} & \textbf{99.9} \\
\bottomrule
\end{tabular}
\end{table}

Table~\ref{tab:comparison} shows that DeepBiESN matches or surpasses the reported recognition rates across most noise conditions, reaching an RR over 99\%  from 15 dB upwards. Notably, at 5 dB, our model achieves the highest RR among all compared methods, demonstrating that 
untrained reservoir dynamics can perform well even in the most challenging scenario. Moreover, it is worth noting that for the highest SNR ($\geq 20$ dB), our model reaches near-perfect recognition rates, closely matching the performance of fully trained models.

\subsection{Comparison with fully trained models}
\label{sec:comparison}

\begin{figure}[t]
    \includegraphics[width=0.9\linewidth]{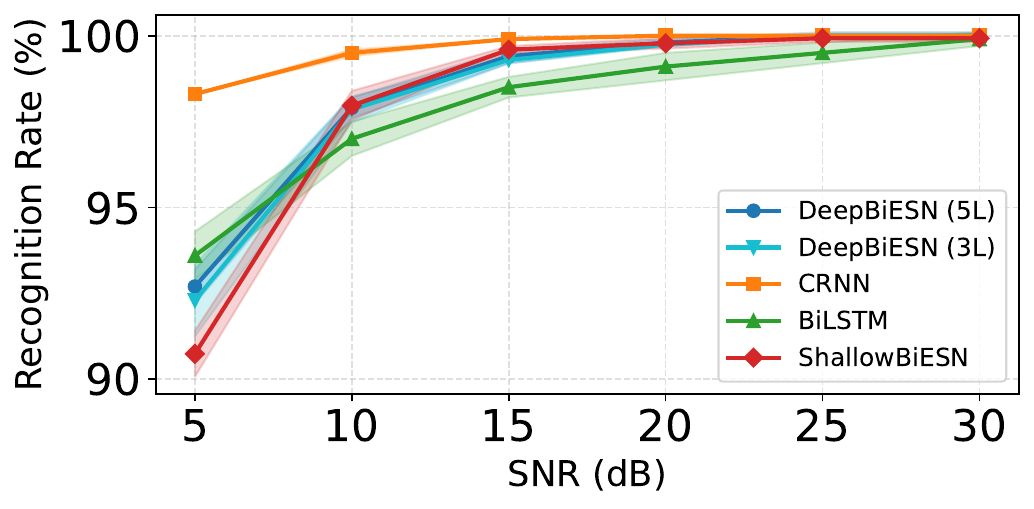}
    \caption{Recognition Rate achieved by the three variants of DeepBiESN (in blue with 5 layers, in cyan with 3, and shallow in red as an ablation on depth), CRNN (in orange), and BiLSTM (in green) at various SNR levels.}
    \label{fig:rr-snr}
\end{figure}

We further compare DeepBiESN against state-of-the-art models that employ a recurrent fully-trained architecture for audio classification, namely a Convolutional Recurrent Neural Network (CRNN) and Bidirectional Long Short-Term Memory (BiLSTM). Our CRNN is based on ~\cite{crnn} and composed of a convolutional front-end for spatial feature extraction, followed by two bidirectional GRU layers that model temporal dependencies through trainable recurrent dynamics.
As for the BiLSTM, we employed the same as that used in~\cite{BiLSTM}. Fig.~\ref{fig:rr-snr} reports the average results in terms of RR across ten runs, obtained with the best configuration of DeepBiESN, comparing the corresponding CRNN and BiLSTM performance.

\begin{figure}[t]
    \includegraphics[width=\linewidth]{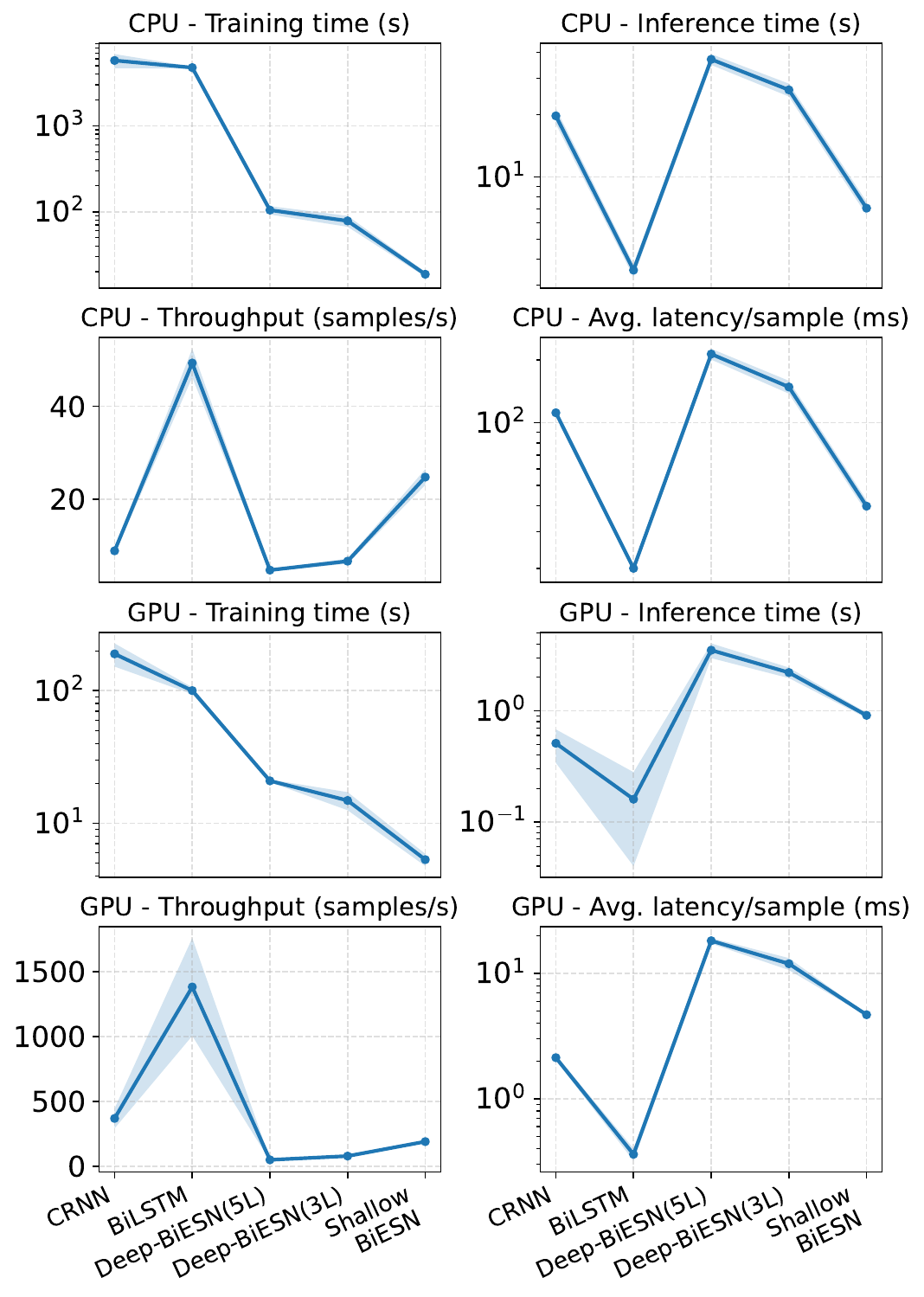}
    \caption{Metrics of efficiency calculated on CPU and GPU on a server platform for all DeepBiESN variants, CRNN, and BiLSTM.}
    \label{fig:cpu}
\end{figure}

\begin{figure}[t]
\centering
    \includegraphics[width=\linewidth]{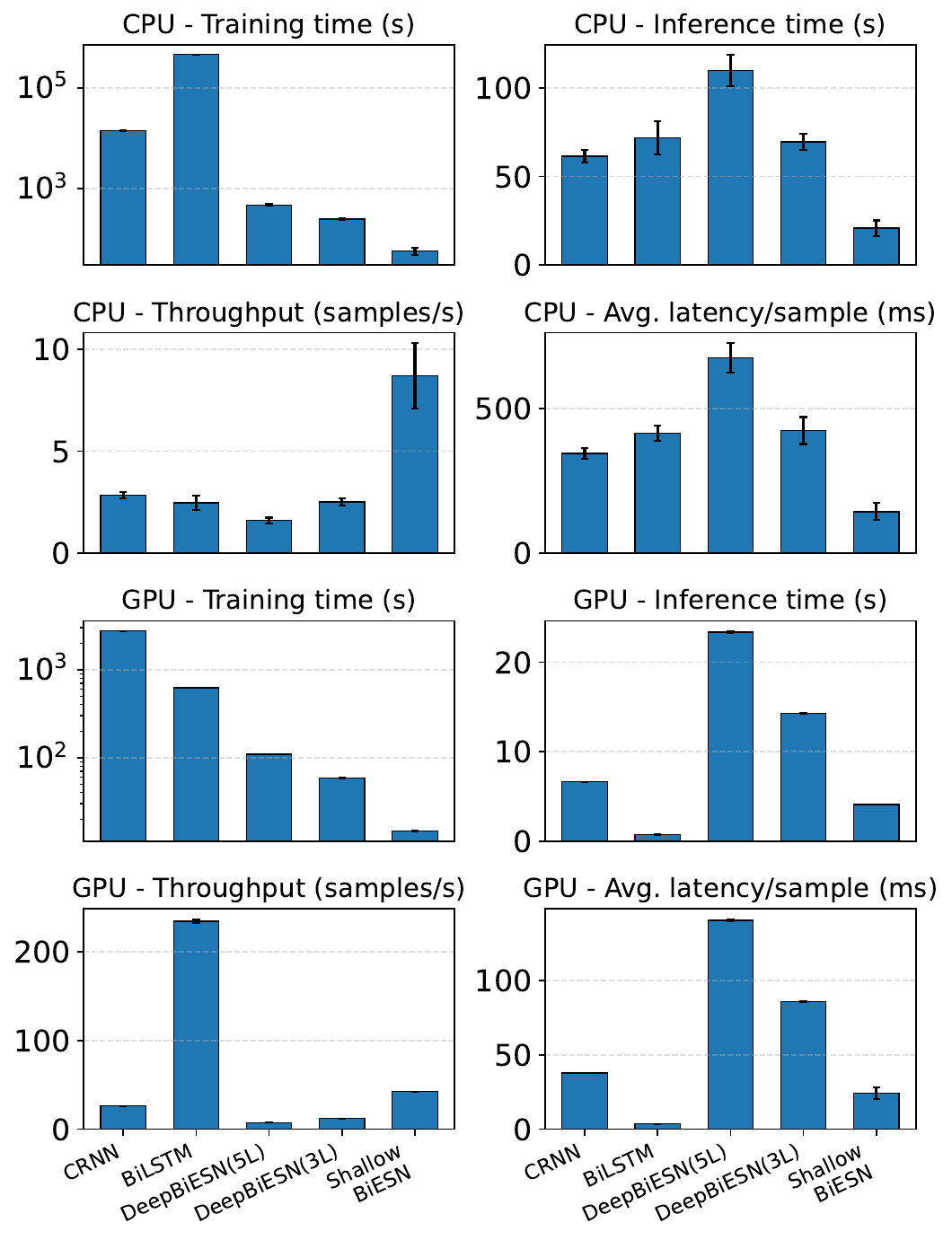}
    \caption{Efficiency metrics measured on the CPU and GPU of the NVIDIA Orin edge device for all DeepBiESN variants, CRNN, and BiLSTM.}
    \label{fig:orin}
\end{figure}

These results highlight two complementary findings. First, compared with fully trained recurrent architectures, DeepBiESN variants remain highly competitive from medium to high SNR conditions, while requiring a substantially cheaper training procedure. Second, the comparison among reservoir architectures with different depths shows that depth plays a meaningful role in the most challenging acoustic scenario: at 5 dB, the deeper configuration performs better than the shallower ones, suggesting that hierarchical reservoir dynamics improve robustness under severe noise. By contrast, at 10 dB and above, the performance gap across deep and shallow variants becomes much smaller, indicating that lighter reservoir models may already provide an effective solution when the acoustic conditions are less adverse.
%
%
We also report an overview of the efficiency of all models in Fig.~\ref{fig:cpu} on the server platform, while in Fig.~\ref{fig:orin} we report the same analysis on the edge NVIDIA Orin device. For efficiency measurements, all models were evaluated in FP32 precision with a batch size of 64 and gradient computation disabled. The first five mini-batches were used for warm-up and excluded from the measurements. The default PyTorch CPU threading configuration was used, resulting in 128 CPU threads on the server platform and 12 CPU threads on the NVIDIA Jetson AGX Orin, matching the number of available physical CPU cores on each device. GPU timings were synchronized using \texttt{torch.cuda.synchronize()} before and after each timed region. Audio feature extraction was performed offline during preprocessing and is therefore not included in the reported timings. Reported results are given as mean $\pm$ standard deviation over ten independent random reinitializations with different seeds. 
Fig.~\ref{fig:cpu} and Fig.~\ref{fig:orin}  show that all DeepBiESN configurations exhibit substantially lower training times compared to BiLSTM and, in particular, CRNN. This advantage is especially pronounced on the CPU settings, where DeepBiESN reduces training time by multiple orders of magnitude. Focusing on inference efficiency on NVIDIA Orin in Fig.~\ref{fig:orin}, on CPU the shallow DeepBiESN configuration achieves the best trade-off: it provides higher throughput, lower per-sample latency, and lower total inference time than other models, while on a server in Fig.~\ref{fig:cpu} is second only to BiLSTM in raw efficiency metrics. Nonetheless, when jointly considering Fig.~\ref{fig:rr-snr}, Fig.~\ref{fig:orin} and Fig.~\ref{fig:cpu}, both the shallow and all deeper DeepBiESN configurations consistently achieve higher accuracy than BiLSTM for SNR values $\geq10$ dB, thus offering a more favorable balance between performance and efficiency. On GPU, DeepBiESN configurations exhibit comparatively worse inference times. This is mainly due to the limited parallelization of our current implementation, whereas CRNN and BiLSTM benefit from highly optimized PyTorch kernels. However, Fig.~\ref{fig:orin} confirms that the shallow DeepBiESN emerges as the most effective trade-off, combining competitive accuracy with superior efficiency in terms of latency, throughput, and inference time.

\section{Robustness to feature representation}
\label{robustness}
\begin{figure}[t]
    \includegraphics[width=\linewidth]{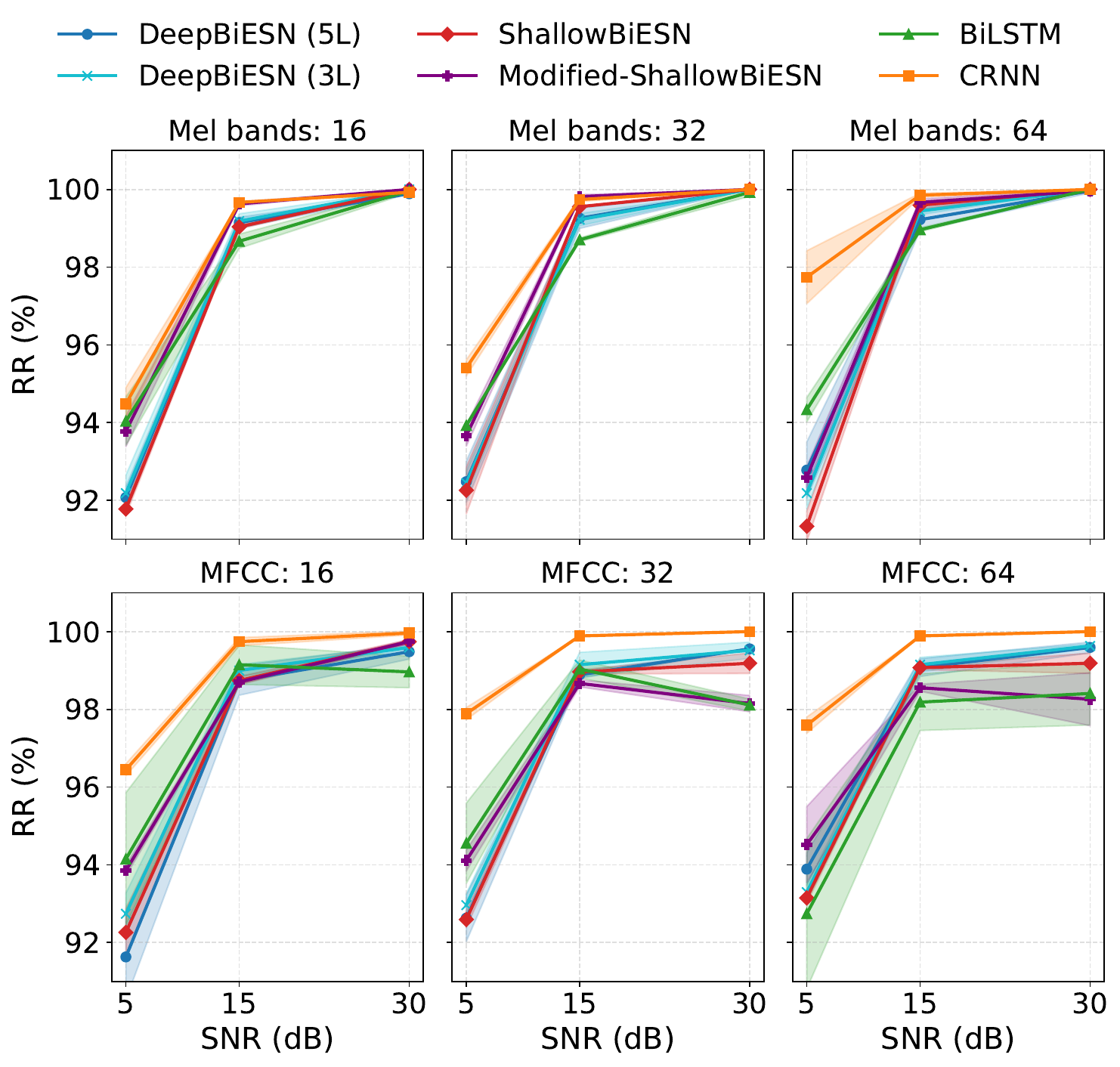}
    \caption{Recognition Rate achieved by the three variants of DeepBiESN, CRNN, and BiLSTM at low, mid, and higher SNR levels, including  Modified-ShallowBiESN as a special case.}
    \label{fig:robustness}
\end{figure}

In this Section, we answer RQ3, investigating the robustness of DeepBiESN \textit{i}) varying the number of Mel filterbanks $M \in \{16,32,64\}$ for the original Mel Spectrogram representation; \textit{ii}) employing Mel-Frequency Cepstral Coefficients (MFCCs) as an alternative feature extraction strategy, experimenting on the same scale as the Mel filterbanks, to assess how the number of coefficients affects recognition performance across different noise conditions. 
Moreover, an ablation study identified $\rho=0.5$ and $a=0.9$ as the optimal setting for the ShallowBiESN configuration, previously highlighted in Section~\ref{sec:comparison} as the best trade-off between efficiency and performance. Accordingly, a modified shallow variant is also considered. Fig.~\ref{fig:robustness} reports the recognition rate (RR) across three SNR conditions—low (5 dB), medium (15 dB), and high (30 dB)—averaged over five runs, including the modified shallow model.
%
As illustrated in Fig.~\ref{fig:robustness}, DeepBiESN consistently achieves high RRs across both Mel and MFCC configurations. Despite the typically greater sensitivity to higher noise of MFCC features, DeepBiESN variants perform on par with CRNN (for SNR$\geq15$), while demonstrating greater stability than BiLSTM, as reflected by lower variance, and even surpassing it in the MFCC$=64$ setting.
Overall, these results suggest that the random reservoir dynamics, once tuned for a specific task, generalize well across different feature representations at various resolutions.

\section{Conclusions and Future Work}
\label{conclusions}
In this work, we applied untrained deep bidirectional recurrent models (DeepBiESN) from the Reservoir Computing (RC) paradigm to audio surveillance on a server platform and on an edge NVIDIA Orin device, using the MIVIA audio events dataset. Experimental results demonstrate that DeepBiESN configurations reach competitive performance in moderate-to-high SNR conditions with significantly less training time than fully trained models, while depth plays a key role in mitigating performance degradation under severe noise. Moreover, the model also exhibits high robustness across various feature representations,  
a desirable property in practical scenarios where the feature extraction pipeline may vary. 
Overall, this study shows that untrained bidirectional reservoir models are a viable solution for audio surveillance in edge scenarios, especially when low training cost and deployment efficiency are key requirements. Our results indicate that architectural depth is a central factor in this trade-off: deeper DeepBiESN variants are more robust in severely noisy conditions, whereas shallower configurations offer a particularly attractive efficiency profile while remaining competitive in moderate and clean acoustic scenarios. This makes the proposed family of models flexible for different operational constraints, rather than tied to a single best configuration. Since the present evaluation is limited to MIVIA Audio Events, the conclusions should be interpreted within this surveillance benchmark, while broader validation on additional environmental and emergency-sound datasets remains necessary. Future work will extend this analysis by evaluating the generalization capabilities of the reservoir approach on additional datasets, comparing it with pretrained and Transformer-based architectures, and conducting further deployment-oriented studies on resource-constrained devices, including Raspberry Pi boards and ultra-low-power microcontroller platforms within the TinyML paradigm.



\balance 

\bibliographystyle{IEEEtran}

\bibliography{bibliography.bib}

\end{document}